\documentclass{article}

\usepackage{microtype}
\usepackage{graphicx}
\usepackage{subfigure}
\usepackage{booktabs}
\usepackage{amsmath,amssymb}
\usepackage{xcolor}
\usepackage{multirow}
\usepackage{tabularx}
\usepackage{enumitem}
\usepackage{siunitx}
\usepackage[hyphens]{url}
\usepackage{hyperref}
\usepackage[capitalise,noabbrev]{cleveref}

\usepackage[accepted]{mlsys2025}

\newcommand{\tkmega}{TK}
\newcommand{\vdcores}{VDCores}
\newcommand{\cudagraph}{CUDA Graph}
\newcommand{\para}[1]{\vspace{2pt}\noindent\textbf{#1.}}
\newcommand{\us}{\si{\micro\second}}

\mlsystitlerunning{Understanding GPU Megakernels}

\makeatletter
\renewcommand{\Notice@String}{}
\makeatother
\hypersetup{pdftitle={Understanding GPU Megakernels: Why Residency Can Mislead},pdfauthor={Xinwei Qiang, Ziyi Xu, Chang Chen, Zaifeng Pan, Zhongkai Yu, Yue Guan, Yufei Ding},pdfsubject={}}

\begin{document}

\twocolumn[
\mlsystitle{Understanding GPU Megakernels: Why Residency Can Mislead}

\begin{mlsysauthorlist}
\mlsysauthor{Xinwei Qiang}{ucsd}
\mlsysauthor{Ziyi Xu}{radix}
\mlsysauthor{Chang Chen}{ucsd}
\mlsysauthor{Zaifeng Pan}{ucsd}
\mlsysauthor{Zhongkai Yu}{ucsd}
\mlsysauthor{Yue Guan}{ucsd}
\mlsysauthor{Yufei Ding}{ucsd}
\end{mlsysauthorlist}
\mlsysaffiliation{ucsd}{University of California, San Diego}
\mlsysaffiliation{radix}{RadixArk}

\mlsyskeywords{Megakernels, GPU, LLM Inference, CUDA Graphs, Programmatic Dependent Launch}

\vskip 0.3in

\begin{abstract}
Megakernels let one persistent GPU program prepare dependent operators
early and execute ready tasks together. We study the benefits and costs of
these capabilities by reconstructing two LLM megakernels,
ThunderKittens (TK) and VDCores, as separate kernels on H100 using their
original operator code.
Our results show that CTA residency can misrepresent opportunities for
useful overlap in both directions.
By overlapping startup and operand loading with programmatic dependent
launch (PDL), the reconstructions reduce decode-step latency by 5.7\% in TK
and achieve comparable performance in VDCores when both executions enforce
the required memory ordering.
Under whole-operator dependencies, TK's persistent execution still gains
more from prefetch, but runtime and completion-signal ordering costs
outweigh this advantage.
Both startup overlap and prefetch depend on resource availability.
Persistent CTAs can reuse storage before retiring, whereas separate
successors must acquire their full CTA footprints before initializing or
loading operands. This admission constraint in both reconstructions
motivates finer resource sharing across kernels. Beyond early
preparation, joint execution requires ready work and spare capacity in the
bottleneck resource. Reserving SMs for
communication can add GEMM waves that erase overlap gains, while filling
idle SMs offers little benefit once memory bandwidth is saturated.
These findings clarify benefits and costs of megakernels and
motivate GPU support for earlier resource reuse across kernels.
\end{abstract}

]

\printAffiliationsAndNotice{}

\section{Introduction}
\label{sec:intro}

Efficient GPU execution is central to low-latency LLM decoding, where
tokens are generated sequentially. Conventional inference systems compose a model from GPU
kernels optimized for its individual operators~\citep{sglang,vllm}.
In small-batch decoding, short operator executions make startup and data
preparation between kernels a significant part of latency.

Recent work explores a different organization: \emph{megakernels}, which
execute multiple operators within a single persistent GPU
grid~\citep{mpk,jin2026event,triton-distributed}. Keeping a GPU program
running across operators can reduce repeated startup, load successor
operands early, and schedule ready work together. ThunderKittens' Llama
megakernel (TK)~\citep{spector2025nobubbles} uses a GPU-resident interpreter
to execute a static task schedule, while
VDCores~\citep{he2026vdcoresresourcedecoupledprogramming} decouples memory
and compute execution to overlap their work.

\begin{figure*}[t]
  \centering
  \includegraphics{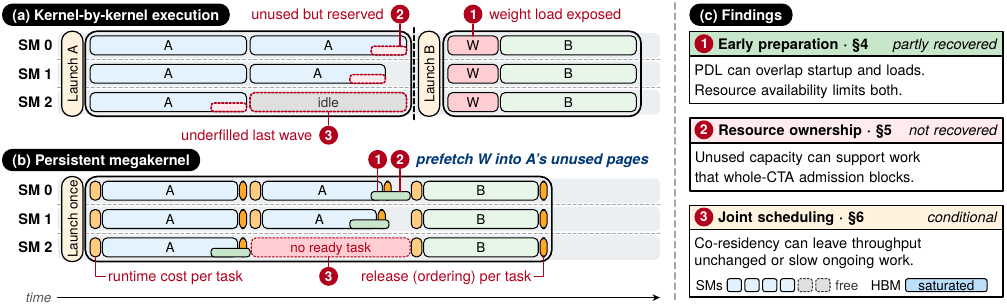}
  \caption{How execution organization changes opportunities for useful overlap.
  Separate kernels retain storage until CTA retirement; persistent CTAs
  reuse it for prefetch but incur runtime and ordering costs.
  Red numbers connect mechanisms in (a,b) to findings in (c).
  Pink marks exposed time or unused capacity; light and dark orange mark
  runtime costs and completion-signal releases. A/B are operators; W is a weight load.}
  \label{fig:overview}
\end{figure*}

What does megakernel execution contribute to inference performance, and when is it preferable to separate kernels? 
End-to-end comparisons conflate operator implementations, synchronization protocols, and execution organization. 
We seek to isolate benefits and costs of execution organization and determine which benefits separate kernels can recover.

We conduct an attribution study of execution organization in small-batch
LLM decoding on H100. We reconstruct TK as separate kernels using its
original operator code, with VDCores as a complementary case study.
We vary successor launch and operand-loading
times to distinguish startup overlap from prefetch benefits. Traces show when
preparation begins and what delays it. We further compare joint and separate
execution to identify when overlapping computation with communication or
combining ready tasks reduces latency, and when resource-sharing and
coordination costs outweigh the gains.

This study reveals a key insight: \emph{CTA residency can misrepresent
opportunities for useful overlap in both directions}. Separate kernels can
be blocked from preparing early despite unused capacity, while allowing
tasks to execute together need not improve throughput (\cref{fig:overview}).

In our reconstructions, programmatic dependent launch (PDL) recovers
opportunities for early preparation by allowing successor kernels to
initialize and load operands before their producers finish. After
correcting completion-signal ordering in the persistent implementations,
the separate-kernel reconstructions reduce decode-step latency by 5.7\%
in TK and achieve comparable performance in VDCores. Our ablations
nevertheless show that TK's persistent execution gains more from prefetch,
although its higher runtime and synchronization costs outweigh that advantage.

TK's larger prefetch benefit motivates examining what limits early
preparation across separate kernels. A persistent CTA can load successor
weights into unused storage while the producer is still computing.
A separate successor, however, must acquire its full CTA footprint before
it can begin preparing. Whole-CTA admission can thus block the loader even
when its warps and first weight buffer could fit beside the producer.
These constraints motivate GPU interfaces that let preparation acquire
resources before a full CTA fits, while preserving efficient operator code.

Beyond preparing dependent operators early, joint scheduling can coordinate
work that is ready to run concurrently. We study two such
opportunities: overlapping computation with communication and combining
ready tiles from independent tasks to fill partial waves. In compute--communication
overlap, reserving SMs for communication can force GEMM into additional
waves, offsetting the overlap. Worker traces show that idle workers can
lack ready tasks. Even with ready work, partial-wave CTAs can saturate
bandwidth, so more resident CTAs need not increase useful throughput.

Our main contributions are:
\begin{itemize}[leftmargin=*]
\item Reconstructions of two megakernels, with controlled experiments in TK
that distinguish startup and prefetch benefits from execution costs,
including the cost of ordering completion signals
(\cref{sec:reconstruction}).
\item A diagnosis of how whole-CTA admission blocks useful preparation
despite unused producer capacity, motivating interfaces for resource reuse
before CTA retirement (\cref{sec:residual}).
\item An analysis of when co-resident execution improves performance, and when
dependencies, extra compute waves, or bandwidth saturation leave throughput
unchanged or slow computation
(\cref{sec:when-fusion-helps,sec:regime-map}).
\end{itemize}

\section{Background and Scope}
\label{sec:background}
\label{subsec:background}

\subsection{GPU Execution and Operator Fusion}
An autoregressive decode step executes a dependency graph of \emph{operators},
including projections, normalization, attention, and collectives.
An ordinary kernel launches a
\emph{grid} of cooperative thread arrays (CTAs), which the GPU places on its
streaming multiprocessors (SMs). A \emph{task} or \emph{tile} is a finer unit
of work. Stream-ordered kernels normally release a successor after its
producer grid completes. A resident megakernel instead executes tasks from
several operators inside one grid.

Fusion can change both scheduling and data movement. MegaMoE, for example,
overlaps expert computation with communication and removes intermediate
buffers~\citep{deepgemm}. Our mechanism ablations preserve operator
computation and compulsory data movement to isolate scheduling effects.

\subsection{Persistent Execution}
Megakernel CTAs execute tasks across operators using static schedules
(TK~\citep{spector2025nobubbles}), dynamic schedules
(MPK~\citep{mpk}), or both (Event Tensor~\citep{jin2026event}).
We examine four capabilities of this organization.

\para{No kernel boundaries}
A decode step runs as one launch, avoiding repeated grid startup and
initialization (\cref{sec:reconstruction}).

\para{Cross-operator prefetch}
TK and MPK reuse shared-memory pages across tasks, loading successor
weights while other tasks still compute. Hopper's Tensor Memory Accelerator
(TMA) performs these loads asynchronously. \vdcores{} uses a resident
memory core that runs ahead of its compute
core~\citep{he2026vdcoresresourcedecoupledprogramming}
(\cref{sec:reconstruction,sec:residual}).

\para{Fine-grained dependencies}
Tasks signal completion through counters in global memory, so a consumer task
can start once the tiles it reads are complete rather than after the whole
producer operator (\cref{subsec:packing}).

\para{Joint scheduling}
A runtime that sees tasks from several operators can run independent work
together, for example by overlapping computation with communication or by
filling SMs that one operator leaves idle~\citep{triton-distributed,mpk}
(\cref{sec:when-fusion-helps}).

\subsection{Cross-Kernel Execution Mechanisms}
\label{subsec:overview}
\begin{figure}[t]
  \centering
  \includegraphics{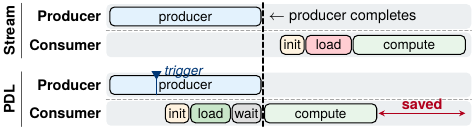}
  \caption{PDL overlaps consumer initialization and independent loads with
  producer execution. Before consuming dependent inputs,
  \texttt{griddepcontrol.wait} waits for producer completion and result
  visibility.}
  \label{fig:pdl-basics}
\end{figure}

CUDA Graphs replay a recorded dependency graph,
amortizing host submission while retaining separate
grids~\citep{nvidia_cuda_graphs}. Programmatic dependent launch (PDL) permits
a producer to trigger its successor before retirement~\citep{cuda-pdl}
(\cref{fig:pdl-basics}).
However, eligibility does not guarantee immediate admission: successor CTAs
still need available SM resources. Once admitted, a successor can load weights
and initialize its pipeline before waiting for the producer's output.

Graphs and PDL retain separate grids: producer registers and shared memory
remain reserved until CTA retirement, and neither creates a
shared task pool across operators.

\subsection{Memory Ordering and Completion Signals}
\label{subsec:memory-ordering}
GPU global memory is coherent at L2, but the SMs' L1 caches are not mutually
coherent. Seeing an updated completion counter therefore does not by itself
ensure that reads see the producer's data. Release/acquire
synchronization orders the producer's writes before publishing completion
and the consumer's dependent reads after observing it, at a scope covering
both CTAs~\citep{nvidia-ptx-isa-93}.

In our H100 binaries, a GPU-scope release update adds
\texttt{MEMBAR.ALL.GPU} before the counter update. The reader polls with
volatile loads that bypass L1, then executes one
\texttt{fence.acquire.gpu} after observing completion. This fence compiles
to \texttt{CCTL.IVALL}, invalidating local L1 cache lines.
The producer barrier orders outstanding accesses
before the signal; the consumer invalidation prevents later L1-cached
reads from reusing stale data. Volatile polling alone lacks this
acquire-side invalidation. Bypassing L1 for the flag alone
neither refreshes cached payload data nor orders the producer's writes.

Across kernels, PDL's launch trigger permits early execution but does not
establish result visibility. Before reading dependent inputs, a successor
can use \texttt{griddepcontrol.wait} to wait for producer completion and result
visibility~\citep{cuda-pdl}. A persistent grid instead needs an explicit
in-grid synchronization protocol for data exchanged between CTAs.

\section{Experimental Approach}
\label{sec:methodology}

To separate scheduling effects from operator implementation, we reconstruct
megakernels with their original operator code as separate kernels. We vary
when successors launch and load operands, when producers trigger them,
and how finely results are released.

\para{Setup}
Single-GPU experiments use an H100 SXM5 GPU (132 SMs), and
GEMM--AllReduce uses eight H100 GPUs. \tkmega{} decodes
Llama-3.2-1B~\citep{grattafiori2024llama3herdmodels} in
BF16 at batch size one, with a 32-token prompt and 128 generated tokens.
For the main \tkmega{} results
(\cref{fig:recovery-controls,fig:prefetch-execution}),
we report the latency of a complete decode step averaged over decode positions.
Every configuration, including the persistent megakernel,
replays one \cudagraph{} per position.

\para{Controls}
Most TK mechanism ablations keep the operator code and each CTA's register and
shared-memory footprint, switch one mechanism at compile time, and differ
from the original compiled code (SASS) only at the targeted sites. Each must
reproduce the persistent kernel's hidden states, KV-cache entries, and
logits within per-tensor tolerances. Joint-scheduling studies instead
compare complete schedules that are each tuned independently.

\para{Completion-signal ordering}
The original TK and VDCores persistent implementations use relaxed
atomic updates and volatile polling for cross-CTA completion signals,
without the ordering described in \cref{subsec:memory-ordering}.
A standalone test reproducing TK's signaling sequence exhibits stale reads.
We use GPU-scope release updates and an acquire fence after each successful
polling loop, labeling these versions
\emph{Ordered persistent}. The main separate-kernel reconstructions use
\texttt{griddepcontrol.wait}. Primary comparisons use these ordered versions.
The original persistent versions serve only as labeled optimistic references
that omit this ordering cost. We assess protocol ordering through source
and SASS inspection, separately from numerical correctness checks,
because passing output tests alone does not establish ordering guarantees.

\para{Timing and traces}
Each TK timing process measures a configuration and the persistent kernel
over 14 timed generations after 3 warm-ups. We rotate configuration order
across five rounds and report medians. All configurations within a round
use the same GPU, with one round on each of five GPUs.
VDCores uses seven interleaved timings per round, each averaging 20 graph
replays.

The compute-bound MLP experiment (\cref{subsec:packing}) uses five rounds
with seven samples per round. Each sample replays one graph of 20 MLP
steps, including counter resets; we report median within-round latency
differences per step.

Timestamp traces come from separate instrumented runs to explain execution
order and overlap; reported performance comes from uninstrumented timing runs.

\section{Recovering Early Preparation Across Kernels}
\label{sec:reconstruction}

We first split TK into separate kernels. We then distinguish the startup
time hidden by an early launch from the additional benefit of prefetching
operands before the preceding operator finishes, and compare prefetch
benefits and execution costs in TK.

\subsection{Launch Overhead and Lost Prefetch}

\tkmega{}~\citep{spector2025nobubbles} executes the sixteen layers and LM head
of Llama-3.2-1B within a single persistent kernel, whose loader warps fetch
the next operator's weights while its compute warps finish the current one.

We extract a separate kernel for each operator. Each layer contributes five
kernels: RMSNorm with QKV projection and RoPE, attention, output projection,
RMSNorm with up/gate projection and SwiGLU, and down projection. These
kernels execute directly, without the persistent interpreter, in order on a
single CUDA stream. With standard CUDA Graph dependencies, each successor
starts only after its predecessor grid completes.

\begin{figure}[ht]
  \centering
  \includegraphics{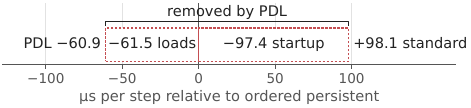}
  \caption{TK separate-kernel latency relative to ordered persistent
  execution with the original dependency granularity.
  Labels show median paired differences.}
  \label{fig:recovery-controls}
\end{figure}

This split execution is about 98.1\,\us{} slower per step than the ordered
version of the original persistent implementation
(\cref{fig:recovery-controls}). Part of this slowdown comes from
the startup gaps that remain between kernels even with graph replay.
In-kernel timestamps show an all-SM idle interval of about 1\,\us{} between
the predecessor's last CTA exit and the successor's first CTA entry.
Across the step's 80 boundaries, these gaps add up to 50--100\,\us{}.

Once the successor enters, it must still fetch weights or KV-cache data
before computing, exposing memory latency as well as startup. Launching
the successor earlier could therefore hide startup alone or also overlap
these loads with the producer's execution. We next separate these two
effects within the split execution.

\subsection{PDL Overlaps Startup and Operand Loading}
\label{subsec:constructive}
\label{subsec:rolling-tail}

PDL lets a successor kernel start and prefetch weights or KV-cache data
before its predecessor finishes. We enable PDL between successive graph
nodes, and each producer triggers its successor after issuing its final
weight load. The successor can then initialize and prefetch, but must wait
for the producer's results before consuming dependent inputs.

With PDL, the separate-kernel reconstruction is 60.9\,\us{} faster per step
than ordered persistent execution, a 5.7\% latency reduction
(\cref{fig:recovery-controls}). It remains 3.4\% slower than the original
persistent implementation, our optimistic reference. The benefit
of PDL does not yet distinguish how much comes from earlier startup and
how much from prefetch.

To separate these effects, we keep PDL enabled but move independent
weight and KV-cache loads after the wait for producer results. CTAs can still enter
early and initialize their pipelines. This configuration remains about
97.4\,\us{} faster than standard graph dependencies, showing the startup
time hidden by PDL. Startup overlap alone brings separate kernels to parity
with ordered persistent execution. Restoring early loads saves a further
61.5\,\us{}.

\begin{figure}[ht]
  \centering
  \includegraphics{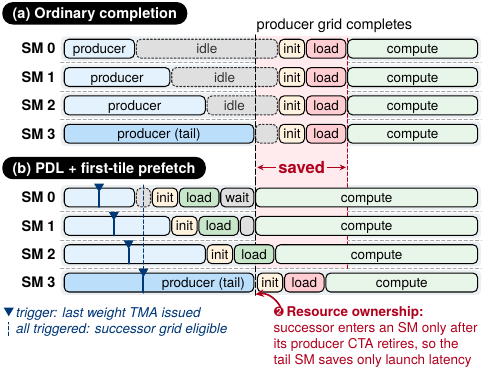}
  \caption{Staggered CTA retirement lets PDL overlap successor startup and
  weight loading with producer execution on other SMs. Dependent computation
  still waits for producer results.}
  \label{fig:pdl-mechanism}
\end{figure}

This overlap raises a placement question. Each extracted CTA retains the
original operator's resource footprint, which nearly exhausts an SM's
registers and shared memory and prevents producer and successor CTAs from
residing together. Where can the successor run before the producer grid
completes? It turns out that producer CTAs retire at different times across
the device. PDL
admits a successor CTA to an SM as its producer CTA retires, while producer
CTAs continue running elsewhere. We call this a \emph{rolling tail}
(\cref{fig:pdl-mechanism}).

Per-SM traces confirm early successor entry at every boundary and early
weight or KV-cache loads at most boundaries. Dependency waits still return
after producer-grid completion. The overlap comes from preparing successors
on available SMs while producer CTAs run elsewhere, with the result dependency
intact.

This makes the producer's retirement pattern important. TK attention uses
only eight SMs, leaving most of the device available to O-projection:
97\% of O-projection CTAs issue weights before attention finishes. At the
next boundary, the median O-projection CTA retires only about 1\,\us{}
before the grid ends, roughly the time a successor CTA needs from entry to
its first weight load, so only about 1\% of up/gate CTAs issue early.
Independent loads benefit only if an SM becomes
available early enough to hide them. Rolling tails create this window
without requiring producer and successor CTAs to share an SM, but do not
provide the same preparation opportunity as reuse inside a live CTA.

As a complementary case, we reconstruct VDCores as separate kernels using
its original operator code. We replay one complete decode step of the same
Llama-3.2-1B model using VDCores' original configuration. This reconstruction
fuses post-attention RMSNorm
into up/gate and the next RMSNorm into down projection. With PDL and
\texttt{griddepcontrol.wait} before consuming producer results, it comes
within 0.1\% of ordered persistent execution.

\subsection{Why Persistent Execution Is Slower}
\label{subsec:persistent-overhead}

We examine the execution and synchronization costs that offset persistent
execution's early-preparation benefits. To exclude gains from fine-grained
result dependencies, both executions wait for the entire producer before
consuming its output. In the persistent variant, each QKV and up/gate
runtime instruction publishes one completion update after its final output
tile, avoiding redundant per-tile updates. With ordering, this variant is
faster than the fine-grained persistent version used above
(\cref{subsec:packing}).

We first disable prefetch in both executions and retain relaxed signaling
in the persistent variant to measure execution-path differences before
adding the required ordering. This optimistic reference is still about
79\,\us{} slower per step than separate kernels with PDL. Both execute the
same operator code, but separate kernels launch each operator directly,
without the persistent interpreter, task dispatch, or storage allocator.

We then measure the additional cost of ordering completion signals within
the same persistent implementation. Ordering adds 19.0\,\us{} per step
without prefetch but 73.5\,\us{} with prefetch. Loading operands earlier
thus makes synchronization more expensive, even with batched completion
updates. With prefetch enabled, the relaxed persistent variant is
28.9\,\us{} faster than separate kernels with PDL. Adding the required
ordering reverses this comparison: ordered persistent is 43.9\,\us{}
slower.

An H100 microbenchmark helps explain this coupling. Release signaling adds
\texttt{MEMBAR.ALL.GPU} before the completion-counter update
(\cref{subsec:memory-ordering}). We hold the signaling sequence fixed while
loader warps on every SM keep background bulk reads in flight. Raising this
volume from 64 to 192\,KiB per SM increases median release latency from 2.8
to 17.2\,\us{}, although background bandwidth stays near 3.2\,TB/s. Relaxed
signaling stays near 0.03\,\us{}. When only the signaling SM issues bulk
reads, release latency stays near 0.6\,\us{} even at 192\,KiB. Release
latency therefore depends on device-wide in-flight traffic. Based on these
measurements, we hypothesize that once bandwidth saturates, additional reads
lengthen memory-system queues, causing release to wait behind this traffic.
Prefetch traffic can thus delay the publication of producer
completion and offset part of the benefit of loading operands early.

\begin{figure}
  \centering
  \includegraphics{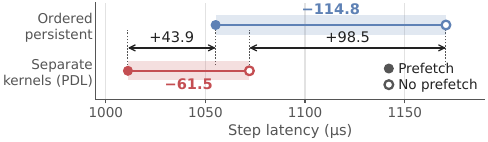}
  \caption{TK prefetch benefits and execution costs. Both executions use
  whole-operator dependencies and required memory ordering.
  Persistent execution batches completion updates and gains more from
  prefetch but remains slower.
  Labels show median paired differences in \us{}.}
  \label{fig:prefetch-execution}
\end{figure}

Despite these costs, persistent execution retains a larger net benefit
from prefetch. With whole-operator dependencies and required ordering,
disabling prefetch increases its latency by 114.8\,\us{}, compared with
61.5\,\us{} for separate kernels with PDL (\cref{fig:prefetch-execution}).
This motivates examining how persistent execution enables earlier
preparation and what prevents separate kernels from accessing the same
opportunities (\cref{sec:residual}).

\section{Resource Ownership Constrains Early Work}
\label{sec:residual}

The split schedule relies on producer CTAs returning resources before a
successor can enter. This admission constraint can expose both startup and
first operand loads, whereas a persistent CTA can prepare the next operator
using resources it already holds. We examine whether separate kernels can
access those resources sooner while preserving operator efficiency.

\subsection{Unused Capacity Can Still Block a Successor}
\label{subsec:admission}

\begin{figure}[t]
  \centering
  \includegraphics{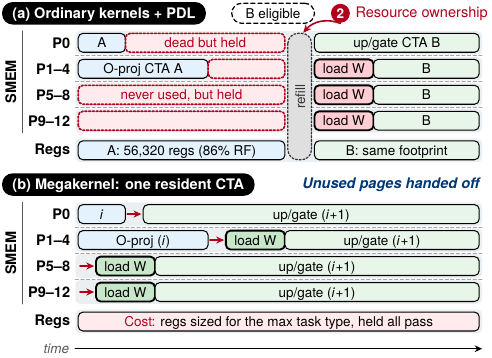}
  \caption{Capacity reuse at O-projection$\rightarrow$up/gate on one SM.
  O-projection uses pages P0--P4 of a thirteen-page reservation.
  (a)~Separate kernels retain the reservation until CTA retirement.
  (b)~A persistent CTA prefetches up/gate weights into unused pages,
  then reuses consumed buffers as they become available.}
  \label{fig:resource-lifetime}
\end{figure}

In TK's persistent megakernel, up/gate weight prefetch can begin before
O-projection completes. Each resident CTA reserves thirteen 16\,KiB
shared-memory pages, but O-projection uses only five. The CTA's loader can
therefore place up/gate's next weight tile in the remaining pages while
O-projection is still running (\cref{fig:resource-lifetime}). Traces of the
persistent kernel show that the first up/gate weight load issues a median
11.5\,\us{} before O-projection completes.

With separate kernels and PDL, however, up/gate must obtain its full CTA resource footprint
before it can prefetch weights. Its shared-memory reservation nearly fills
an SM, so even releasing O-projection's unused pages would not leave enough
space for the up/gate CTA alongside O-projection. Registers pose a second
constraint: O-projection alone occupies 86\% of the SM register file, leaving
too few registers for the up/gate CTA. The persistent megakernel can prefetch
up/gate weights into its unused pages without admitting a separate CTA with
both resource requirements. The blocked work is weight loading, but
admission tests the resources needed for the entire compute phase.

We test an alternative explanation for the missing prefetch window:
a few late O-projection CTAs may delay PDL's launch trigger for up/gate.
PDL permits successor admission only after every producer CTA has issued its
trigger~\citep{cuda-pdl}. Attention occupies eight of H100's 132 SMs, so the
O-projection CTAs that enter those SMs cannot load weights and issue their
triggers until attention finishes. These stragglers could therefore block
up/gate admission even when other SMs have capacity. To remove this delay,
we pack O-projection's weight tiles into 124 CTAs that fit on the remaining
SMs. The change has little effect on when up/gate can prefetch. Even with earlier
triggers, O-projection CTAs still wait for attention results before computing
and then finish close together, leaving little capacity for up/gate to enter
early. Late triggers from a few stragglers therefore do not explain most of
the missing prefetch window.

The same admission constraint limits startup overlap in VDCores. Its
attention grid occupies 64 SMs, and traces show that about 47\% of
O-projection CTAs enter only after attention finishes. These CTAs must
initialize queues and barriers and check the input dependency before
issuing activation loads, so this startup remains exposed even with weight
prefetch disabled. The persistent memory core is already resident and
does not repeat this entry path. Resource availability therefore controls
both how much successor startup PDL can hide and when prefetch can begin.

\subsection{Software Recovery Can Add Execution Costs}

The admission constraint suggests reducing CTA footprints so that
O-projection and up/gate can reside on the same SM. Under whole-CTA
admission, however, this requires shrinking the resources reserved for
computation even though only the successor's loader needs to enter early.
Both shared-memory reservations and register demand would have to fall.
These reductions also constrain the operator's compute phase and can
reduce its efficiency. As a diagnostic of this tradeoff, we evaluate a
low-register O-projection that allows two of its own CTAs to share an SM.
It is 3.4\% slower than the original in isolation.
Earlier admission must be weighed
against the execution cost of the smaller-footprint operators.

VDCores shows why an improvement at one boundary need not survive across
the operator sequence. Reducing shared-memory reservations allows
O-projection to prefetch all its weights alongside attention. In instrumented
traces, the O-projection level shortens from about 4.7 to 2.9\,\us{}, yet the
corresponding full step becomes slower. The early O-projection CTAs retain
their reservations, leaving fewer resources for up/gate to enter next.
At other boundaries, smaller reservations remove a weight-buffering stage
and slow the pipelines. In these schedules, gaining early loads at one
boundary can delay entry or reduce pipeline efficiency at another. Earlier
admission changes both the preparation window and the resources
available to other work.

\subsection{Implications for Resource Interfaces}
\label{subsec:handoff}

Early weight prefetch needs fewer resources than full computation. At
TK's O-projection$\rightarrow$up/gate boundary, the loader and its first
weight tile could fit beside the producer even though the full successor
CTA cannot. The interface should admit this preparation first, then let
the successor grow as producer resources are released.

Ordinary inference kernels encounter this constraint. In a GLM-5.2
SGLang configuration, 188 of 822 adjacent same-stream kernel pairs cannot
place one CTA from each kernel on the same SM, mostly because of register
demand~\citep{sglang,deepgemm,flashattention3}.

These cases motivate separating shared-memory, register, and warp lifetimes
from the complete CTA, preserving efficient computation without requiring a
persistent interpreter.

The producer should be able to release unused shared memory and consumed
buffers while retaining live buffers, preserving its pipeline as capacity
becomes available for successor weights.

Reassigning this storage must preserve forward progress. Before transferring
released capacity to the successor, the producer must commit to no further
shared-memory allocation. Otherwise, it could wait for capacity held by a
successor that awaits its results. Blackwell's tensor memory (TMEM) provides
a precedent: \texttt{tcgen05.relinquish\_alloc\_permit} commits a CTA to no
further TMEM allocation~\citep{nvidia-ptx-isa-93}. A shared-memory interface
could similarly separate this commitment from the later release of buffers
that still contain live data.

Shared-memory release alone is insufficient when registers block the loader.
The interface could allow individual producer warps
to retire and return their registers and resident-warp slots. For example,
a loader warp could retire after issuing its final TMA load and completing
any remaining synchronization duties, while the compute warps continue
running. Its shared-memory buffers would remain allocated until
transfers and consumers finish using them. Warp-level dispatch and resource
reclamation, as explored by WarpMan, provide a basis for this
behavior~\citep{warpman}.

Partial CTA admission must then check the loader's resource requirements
rather than the complete CTA's. Its warps can prefetch into available shared
memory while compute warps await resources. As the producer releases the
remaining capacity, the successor can admit compute warps and grow to its
full footprint. \Cref{tab:handoff} summarizes these capabilities.

\begin{table}
  \centering
  \caption{Proposed resource capabilities for early weight prefetch across
  kernels. Their performance remains unmeasured.}
  \label{tab:handoff}
  \footnotesize
  \begin{tabularx}{\columnwidth}{@{}p{0.29\columnwidth}X@{}}
    \toprule
    Capability & Effect on execution \\
    \midrule
    Partial admission and retirement & Completed producer warps release resources so successor loader warps can enter before a full CTA fits. \\
    Phase-scoped release & Producer returns shared-memory buffers after their last use, before its warps retire. \\
    Ordered growth & Successor acquires its remaining resources and admits compute warps as capacity becomes available. \\
    \bottomrule
  \end{tabularx}
\end{table}

However, software alone cannot provide these capabilities, because the
hardware allocates and reclaims resources per CTA.
With identical live CTA footprints, our H100 probes observe admission limits
of 145 versus 96\,KiB under different allocation histories, consistent with
contiguous CTA allocations in a circular shared-memory arena. Under this model,
phase-scoped release requires released address ranges to become independently allocatable,
while live buffers retain their addresses.

Similarly, for partial admission and ordered growth, admission logic must
track changing resource allocations and distinguish
resident warps from those awaiting entry, while preserving CTA
synchronization state.

\section{When Joint Scheduling Adds Value}
\label{sec:when-fusion-helps}

\Cref{sec:reconstruction,sec:residual} focus on preparing a dependent successor
before its producer finishes. The software recovery experiments also show
that earlier admission can delay or slow other work. We now ask when
sharing resources between independent tasks from different operators
improves performance. Communication for
completed output tiles can run alongside computation of later tiles, while
independent compute tasks can fill the idle SMs of a partial wave.
Co-residency does not by itself add execution capacity: ready tasks may
use spare bottleneck resources or compete for capacity needed by ongoing
work.
We examine when the overlap gained outweighs the costs of sharing resources and
coordinating execution.

\subsection{Compute--Communication Overlap}

\begin{figure}
  \centering
  \includegraphics{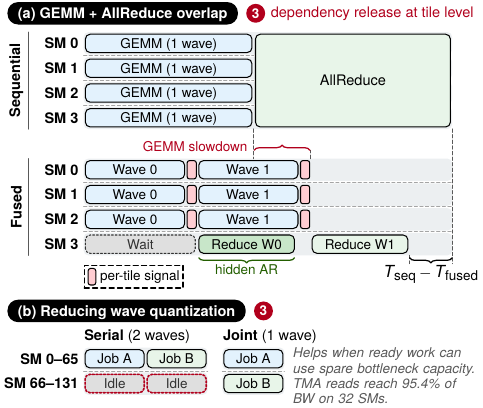}
  \caption{Opportunities for joint scheduling. (a)~Releasing
  GEMM tiles early can hide communication, but reserving resources and
  signaling add cost. (b)~Jointly executing independent tasks may reduce wave quantization when both are ready.}
  \label{fig:overlap-packing}
\end{figure}

Multi-GPU execution adds inter-GPU communication to the local operator
sequence. A tensor-parallel row projection consists of a local GEMM followed
by an AllReduce. A fused schedule can begin communicating completed output tiles
while later GEMM tiles are still being computed
(\cref{fig:overlap-packing}a), exploiting the different resources used by
computation and communication. Prior systems use this approach for fine-grained
compute--communication overlap~\citep{t3-overlap,flux-overlap,triton-distributed}.
MegaMoE applies such overlap to expert-parallel MoE layers, assigning
communication to dedicated warps within each CTA rather than to reserved
SMs~\citep{deepgemm}. However,
communication also needs GPU execution resources, leaving fewer for GEMM.
Does hiding communication outweigh the resulting slowdown in computation?

We evaluate Triton-distributed's fused GEMM--AllReduce
implementation~\citep{triton-distributed}, which assigns some SMs to
communication and the others to GEMM. Without overlap, GEMM can use all SMs before AllReduce begins,
and its best tile configuration may therefore differ. Reusing the fused
GEMM's SM allocation and tile configuration for the sequential baseline
could underestimate sequential performance. We therefore tune both
implementations independently and compare their full GEMM--AllReduce
latencies to determine whether overlap is beneficial.

The experiment covers Qwen3-8B and Qwen3-32B~\citep{yang2025qwen3technicalreport}
attention-output and MLP-down
projections at twelve batch sizes from 1 to 256 with eight-way tensor
parallelism on eight H100 GPUs. The fused candidate tunes its tiles and
GEMM/communication SM split. The sequential candidate tunes GEMM over all 132 SMs and the
following multimem AllReduce independently.

\begin{figure}
  \centering
  \includegraphics{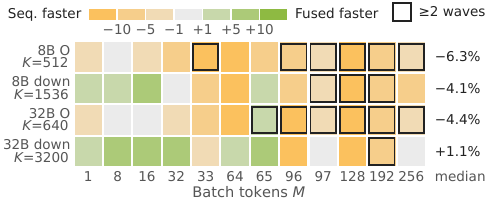}
  \caption{Fused GEMM--AllReduce latency reduction (\%) over independently
  tuned sequential execution for Qwen3-8B/32B O and down projections on
  eight H100s (TP8). $K$ is the local reduction dimension. Outlines mark
  fused GEMMs requiring multiple waves after SM reservation.
  Right: median per projection.}
  \label{fig:overlap-results}
\end{figure}

Across the 48 points, fusion is more than 1\% faster at 11 points, more than
1\% slower at 31, and within 1\% at the remaining six
(\cref{fig:overlap-results}). We first examine whether the GEMM
provides enough computation to hide communication, then measure how much
reserving communication SMs slows that computation.

\para{Short GEMMs leave limited room for overlap}
We estimate the AllReduce time as sequential latency minus GEMM-only
latency. This cross-run estimate includes launch and synchronization costs
and has an approximate uncertainty of $\pm1\,\us{}$ per point.
In these measurements, fusion never wins by more than 1\% at the
24 points where GEMM takes less than about 90\% of this estimated time.
At a given batch size, a model's attention-output and MLP-down projections
produce outputs of the same size and communicate the same amount of data.
The down-projection's larger local reduction dimension supplies more
computation to overlap with that communication. O-projection wins at only
one of 24 points, whereas ten of the eleven wins are down-projections at
batch sizes up to 65.

\para{SM reservation matters at wave boundaries}
A \emph{wave} here is one round of tiles over the SMs assigned to GEMM.
We time GEMM alone with and without the communication SM reservation.
With the same tile configuration, reserving communication SMs changes
isolated GEMM latency by less than 0.3\,\us{} at all 32 single-wave points.
The cost rises when the output tiles no longer fit on the remaining SMs
and require another wave. For Qwen3-8B O-projection at batch 128, both
candidates select the same tile configuration: the 128 output tiles fit on
132 SMs but need a second wave on the fused GEMM's 116. Timed alone, GEMM
slows from 4.9 to 7.2\,\us{}, an increase exceeding the fused schedule's
net loss of 2.2\,\us{}. Fifteen of the sixteen multi-wave configurations are slower
than sequential execution. The remaining configuration still wins, showing
that overlap can sometimes repay the extra GEMM wave.

Seven single-wave points above the time-ratio threshold still lose, so a
long GEMM and a single wave do not guarantee that overlap pays. The gain
from hidden communication must cover the compute slowdown and coordination
costs. Wave boundaries make the compute penalty especially abrupt.

\subsection{Reducing Wave Quantization}
\label{subsec:packing}
\label{subsec:bandwidth}

Joint scheduling can also reduce wave quantization by filling the idle SMs
in an operator's final, partial wave. If another operator's tiles
are ready, they can run alongside that wave rather than in a later wave
of their own (\cref{fig:overlap-packing}b). Two conditions determine whether
this opportunity improves latency: independent work must be ready, and the
idle SMs must provide capacity in the resource that limits throughput.

\para{Dependencies limit ready work}
Fine-grained scheduling can exploit available parallelism, but cannot create
it. Finer dependency release can make successor tiles ready before their
producer finishes. The opportunity depends on whether
consumer readiness coincides with available execution capacity.

Publishing finer-grained readiness also has a cost. In TK, the
whole-operator variant used in \cref{subsec:persistent-overhead} replaces
per-head and per-shard dependencies at QKV$\rightarrow$attention and
up/gate$\rightarrow$down with one completion counter per operator.
This permits each producer runtime instruction to publish a single release
update after its final tile, while retaining the same tile decomposition
and prefetch. With required ordering in both versions, the coarser variant
reduces decode-step latency by 1.6\% (about 17\,\us{}).
In this schedule, the earlier readiness provided by fine-grained
dependencies does not repay the cost of more frequent completion updates.

\begin{figure}
  \centering
  \includegraphics{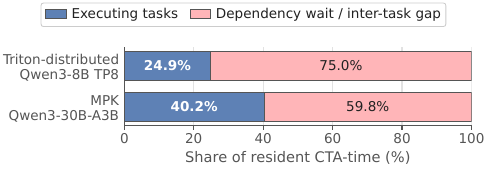}
  \caption{Share of resident worker time spent executing tasks versus waiting
  inside persistent grids. Triton-distributed: Qwen3-8B, TP8, 132 workers.
  MPK: Qwen3-30B-A3B, one GPU, 128 workers. Fractions refer to resident
  worker-time, not GPU utilization.}
  \label{fig:dependency-width}
\end{figure}

In full-model execution, persistent-worker traces show that ready tasks
can remain scarce even with fine-grained scheduling.
In MPK's Qwen3-30B-A3B trace, workers execute tasks for only 40.2\% of their
resident time, spending the rest between tasks (\cref{fig:dependency-width}).
The task graph helps explain these waits: an attention stage has only four
tasks, followed by an O-projection with 32 tasks, for 128 resident workers.
This narrow dependency chain limits how many workers can execute, even
though the runtime can dispatch individual tasks. Triton-distributed shows
similar waiting: its Qwen3-8B TP8 path spends 75.0\% of resident CTA-time
waiting for dependencies.

A larger physical grid does not guarantee more useful work. In a GLM-5.2
trace, decode DeepGEMM launches 132 CTAs for 8--128 logical output tiles,
while a query-length-four FA3 case uses 36 useful scheduler CTAs behind 132
physical CTAs. Joint scheduling needs additional tiles whose dependencies
are satisfied, rather than simply more resident workers.

\begin{figure}[ht]
  \centering
  \includegraphics{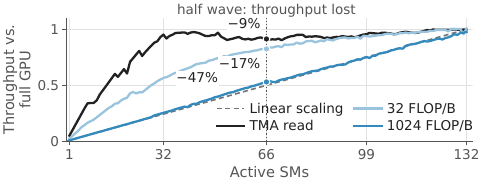}
  \caption{Throughput versus active SMs, normalized to the full GPU, for
  TMA reads and tensor-core computation. Labels show half-wave throughput
  loss. The dashed line indicates linear scaling.}
  \label{fig:bandwidth}
\end{figure}

\para{Bandwidth limits the benefit of additional ready work}
Even when independent tasks are ready, filling idle SMs need not improve
throughput. A bandwidth-bound task may already saturate memory with only
part of the GPU active. Additional memory-bound tasks then compete for the
same bandwidth. In contrast, an underfilled compute-bound wave leaves
arithmetic units available for other compute tasks.
In a steady-state TMA read experiment, 32 of H100's 132 SMs supply 95.4\%
of the full-device bandwidth (\cref{fig:bandwidth}). Filling the remaining
SMs offers little additional bandwidth for this workload.

The cost of leaving SMs idle grows as the workload becomes more
compute-intensive. A kernel that performs arithmetic on its loads needs more
SMs than the read stream to approach full throughput: a half-filled wave
loses about 17\% at 32\,FLOP/B and nearly half at 1024\,FLOP/B
(\cref{fig:bandwidth}).

We construct a compute-bound MLP, with fused SwiGLU and
split-$K$ down projection, in a persistent Triton kernel with one worker
per SM. With 2.12 waves of
up/gate tiles and 1.88 waves of down tiles, releasing completed input slices
lets ready down tiles fill the sparse producer tail, reducing latency by
about 13\% versus whole-operator release.

For a given workload, this opportunity should be evaluated after tuning
each operator's tile configuration.

\section{Implications for Runtimes and GPUs}
\label{sec:regime-map}

These results suggest three rules for choosing and tuning execution forms.

\para{Tune prefetch together with synchronization}
Additional in-flight loads can increase release latency. Tune prefetch with
required ordering enabled, and judge its net benefit by complete-step latency.

\para{Account for the cost of earlier admission}
If earlier admission requires smaller CTAs, weigh the operator slowdown
against the overlap gained. Validate the full sequence: resources held by
an early successor can delay later work.

\para{Check wave boundaries before reserving resources}
Reserving communication SMs can add a GEMM wave that costs more than the
overlap saves. Compare independently tuned joint and separate schedules.
For partial-wave packing, require ready work and spare bottleneck capacity:
idle SMs alone are insufficient.

\section{Related Work}
\label{sec:related}

\para{Persistent execution and graph scheduling}
Persistent threads and task-queue frameworks established device-side work
distribution before LLM inference~\citep{gupta-persistent-threads,whippletree,versapipe}.
Neural-network systems apply residency in several ways. Persistent RNNs and
GRNN retain recurrent weights on chip~\citep{persistent-rnn,grnn}, Deep Voice
executes WaveNet generation in one launch~\citep{deepvoice}, and VPPS and
Cortex support persistent or compiled execution for dynamic and recursive
models~\citep{vpps,cortex}. Rammer exposes tasks across operator boundaries
to a shared scheduler~\citep{rammer}. Cocktailer, Souffle, and MonoNN also
coordinate larger computation graphs~\citep{cocktailer,souffle,mononn}, while
RECom and RecFlex apply composition to recommendation
workloads~\citep{recom,recflex}. These designs coordinate work across
operators within one kernel. Our reconstruction isolates the
schedule effects for a fixed set of operators and data transfers. The
operator-efficiency costs we observe also echo earlier work on divergence
and register pressure in path-tracing megakernels~\citep{laine-megakernels}.

\para{LLM megakernels and cross-kernel execution}
TK, MPK, Event Tensor, VDCores, and Triton-distributed provide static,
dynamic, and resource-decoupled organizations for LLM
execution~\citep{spector2025nobubbles,mpk,jin2026event,he2026vdcoresresourcedecoupledprogramming,triton-distributed}.
Event Tensor compares against engines using Graphs and PDL and
studies pipelining, wave quantization, and prefetch.
Using per-SM traces and prefetch ablations, we explain how separate
kernels recover early preparation. Resource experiments identify opportunities
blocked across kernels.

\para{Dependency release and joint scheduling}
CUDA Graphs and PDL expose graph replay and early dependent
launch~\citep{nvidia_cuda_graphs,cuda-pdl}. Wireframe and BlockMaestro
study related launch and dependency mechanisms~\citep{wireframe,blockmaestro}.
Kernelet co-schedules sliced independent kernels~\citep{kernelet}.
Decomposition, T3, FLUX, TileLink, and Triton-distributed enable fine-grained
compute--communication overlap~\citep{wang-overlap,t3-overlap,flux-overlap,tilelink,triton-distributed}.
DeepGEMM's Mega MoE and mKernel extend coordinated schedules
across distributed operators~\citep{deepgemm,mkernel}, and Stream-K addresses wave quantization
within an operator~\citep{stream-k}. Our measurements examine the compute
schedule left after resources are assigned to overlapping roles, and the
bandwidth and readiness conditions that limit gains from reducing wave
quantization across operators.

\para{On-chip resource management}
WarpMan, register-file virtualization, RegMutex, scratchpad sharing, Zorua,
and Virtual Thread relax ownership or dispatch
granularity~\citep{warpman,rfv,regmutex,smm,jatala-scratchpad,zorua,virtualthread}.
SMK and Warped-Slicer partition SM resources among kernels, while FIBER
separates register ownership from thread execution~\citep{smk,warpedslicer,fiber}.
The resource capabilities discussed here build on this body of work. The
reconstructions motivate a specific dependent-kernel use: admitting a loader
with a small initial footprint, handing over capacity during execution, and
then obtaining the full compute footprint. Evaluating that interface is a
future step beyond the measured constraints.

\section{Conclusion}
\label{sec:conclusion}

Our reconstructions show that separate kernels with PDL outperform ordered
persistent execution in TK and achieve comparable latency in VDCores.
Under whole-operator dependencies, TK's persistent execution obtains a
larger prefetch benefit, but runtime and completion-signal ordering costs
outweigh this advantage. Joint scheduling can use idle compute capacity, but resource sharing
can add compute waves, and idle SMs need not provide spare bandwidth.
Together, these results show how residency can misrepresent useful overlap
in both directions: whole-CTA admission can block preparation that fits in
unused capacity, while co-resident execution can leave throughput unchanged
or slow ongoing work. GPU interfaces should let preparation acquire
resources before a full CTA fits and runtimes must also account for how
sharing those resources changes the execution of other work.
\label{sec:main-end}

\bibliography{refs}
\bibliographystyle{mlsys2025}

\section*{AI Usage Disclosure}

We used OpenAI Codex and Anthropic Claude Code to assist with literature
search, experiment and ablation design, code implementation and debugging,
analysis of measured results, and manuscript drafting, revision, and
review. Code assistance covered GPU kernels, benchmarking and profiling
scripts, and data-analysis and plotting scripts. AI assistance was used
iteratively in the reconstruction, resource-admission, and joint-scheduling
studies and in preparing their presentation.

The reported performance results come from executed GPU experiments.
The authors retain responsibility for the experimental
methodology, validation of implementations and measurements,
interpretation of results, accuracy of references, and final manuscript.

\end{document}